\documentclass[prb,twocolumn,superscriptaddress,notitlepage]{revtex4-2}
\usepackage[utf8]{inputenc}
\usepackage[american]{babel}
\usepackage[pdftex]{graphicx}  
\usepackage{dcolumn}
\usepackage{bm}
\usepackage{dsfont}
\usepackage{amsmath,amsthm,amssymb}
\usepackage{hyperref}
\usepackage{xcolor}
\hypersetup{colorlinks,bookmarksopen,bookmarksnumbered,
citecolor=blue,
linkcolor=blue,
pdfstartview=false,
urlcolor=blue}
\usepackage{braket}

\usepackage[normalem]{ulem}
\usepackage{orcidlink}

\begin{document}
 \title{Pauli Spectrum and Non-stabilizerness of Typical Quantum Many-Body States}

\author{Xhek Turkeshi~\orcidlink{0000-0003-1093-3771}}
\email{turkeshi@uni-koeln.de}
\affiliation{Institut f\"ur Theoretische Physik, Universit\"at zu K\"oln, Z\"ulpicher Strasse 77, 50937 K\"oln, Germany}
\author{Anatoly Dymarsky~\orcidlink{0000-0001-5762-6774}}
\affiliation{Department of Physics, University of Kentucky, Lexington, Kentucky, USA, 40506}
\author{Piotr Sierant~\orcidlink{0000-0001-9219-7274}}
\affiliation{ICFO-Institut de Ci\`encies Fot\`oniques, The Barcelona Institute of Science and Technology, Av. Carl Friedrich Gauss 3, 08860 Castelldefels (Barcelona), Spain}
\date{\today}

\begin{abstract}
An important question of quantum information is to characterize genuinely quantum (beyond-Clifford) resources necessary for universal quantum computing. 
Here, we use the Pauli spectrum to quantify how ``magic'', beyond Clifford, typical many-qubit states are. 
We first present a phenomenological picture of the Pauli spectrum based on quantum typicality 
and then confirm it for Haar random states. We then introduce filtered stabilizer entropy, a magic measure that can resolve the difference between typical and atypical states. 
We proceed with the numerical study of the Pauli spectrum of states created by random circuits as well as for eigenstates of chaotic Hamiltonians. We find that in both cases Pauli spectrum approaches the one of Haar random states, up to exponentially suppressed tails. We discuss how the Pauli spectrum changes when ergodicity is broken due to disorder. Our results underscore the difference between typical and atypical states from the point of view of quantum information. 
\end{abstract}
\maketitle

\section{Introduction}
Quantifying non-stabilizerness or how ``magic'' quantum states of extended quantum systems are~\cite{bravyi2005universalquantumcomputation}, is central to quantum information processing and computation, as it determines the amount of beyond-classical  (non-Clifford) operations needed to perform a quantum task~\cite{Kitaev2003, Gottesman1999demonstratingtheviability,Veitch2014theresourcetheory,bravyi2016tradingclassicaland,chitambar2019quantumresourcetheories,liu2022manybodyquantummagic}. Clifford operations are central to many areas of physics, from quantum error correction~\cite{gottesman1998theoryoffaulttolerant,nielsen00,eastin2009restrictions}, to condensed matter theory, including entanglement dynamics~\cite{nahum2017quantum,zhou2019emergent,zhou2020entanglementmembrane,sierant2023membrane,Fritzsch21, Claeys21} and measurement-induced transitions~\cite{li2019measurementdrivenentanglement, fisher2023randomquantumcircuits,potter2022quantumsciencesandtechnology,sierant2022measurementinducedphase,jian2020measurementinducedcriticality,li2021statisticalmechanicsmodel}, but do not by themselves provide quantum advantage~\cite{gottesman1997stabilizer,aaronson2004improvedsimulationof}.

We propose quantifying non-stabilizerness through the statistical properties of the Pauli spectrum (PS). For an $N$-qubit state $|\Psi\rangle$ on the $d=2^N$ dimensional Hilbert space, the PS is defined as the
set of $d^2$ real numbers~\cite{Beverland2020}
\begin{equation}
    \mathrm{spec}(|\Psi\rangle) = \left\lbrace \langle \Psi| P |\Psi\rangle,~ P\in \mathcal{P}_N\right\rbrace.\label{eq:paulispec}
\end{equation}
Here $\mathcal{P}_N=\{ P_1\otimes P_2\otimes \cdots \otimes P_N \}$ are the 
Pauli strings defined via the Pauli matrices $ P_i \in \{X_i,Y_i,Z_i,I_i\}$ ~\footnote{Compared to~\cite{Beverland2020}, we prefer to avoid the absolute value in the definition, allowing for a more precise characterization. Reverting to~\cite{Beverland2020} from ~\eqref{eq:paulispec} is straightforward.}. 
The Pauli spectrum~\eqref{eq:paulispec} is sufficient to compute various non-stabilizerness quantifiers, including nullity~\cite{Beverland2020,jiang2023lowerboundfor,leone2023learning}, the stabilizer rank~\cite{Qassim2021improvedupperbounds}, and the stabilizer R\'enyi entropy (SRE)~\cite{leone2022stabilizerrenyientropy,haug2023efficient,oliviero2022measuringmagicon,turkeshi2023measuring,gu2023little,tirrito2023quantifying}.
The latter quantity can be readily computed with the help of Monte-Carlo sampling for variational wave-functions~\cite{tarabunga2023magic} and for matrix product states~\cite{haug2023quantifyingnonstabilizernessof,haug2023stabilizerentropiesand,lami2023quantum,tarabunga2023manybodymagic,chen2023magic,lami2025}, providing a systematic characterization of non-stabilizerness of the ground state of one-dimensional systems~\cite{oliviero2021transitionsinentanglementcomplexity,oliviero2022magicstateresourcetheory}. 
The Pauli spectrum contains more refined information about non-stabilizerness than the SRE, similar to the relation between the entanglement spectrum and the entanglement R\'enyi entropy~\cite{calabrese2004entanglemententropyand, calabrese2008spectrum, Laflorencie_2016}.

In this work, we study the statistical properties of the Pauli spectrum and evaluate the stabilizer entropy of typical quantum states. After presenting a typicality-based picture~\cite{Goldstein_2006,reimann2007typicality,gemmer2009dynamical,Goold2016},
we introduce \emph{filtered stabilizer entropy}, a non-stabilizerness measure that suitably distinguishes between generic and atypical (low entangled) many-body states.
We then corroborate the overall picture by deriving the exact Pauli spectrum of Haar-random (unitary and orthogonal) quantum states and via extensive numerical simulations for states produced by random circuits and high-energy eigenstates of chaotic spin chains.

\section{Pauli spectrum and stabilizer entropy}
It is convenient to think about the PS~\eqref{eq:paulispec} as a probability distribution
\begin{equation}
\label{P}
    {\Pi(x) = \sum_{x_P\in \mathrm{spec}(|\Psi\rangle)} \delta(x-x_P)/d^2}.
\end{equation} 
In what follows, we refer to \eqref{P} as the Pauli spectrum with a slight notational abuse. The PS can readily distinguish the stabilizer states from the non-stabilizer ones. To illustrate this point, we note 
that the Pauli spectrum of any stabilizer state has exactly $d$ elements equal to ${\pm1}$~\footnote{Either a stabilizer state has $d$ elements +1, or it has $d/2$ elements +1 and $d/2$ elements -1.}, with all others being zero. As we will see shortly, this is very different from a typical quantum state. 

The Pauli spectrum fully determines the
SRE~\cite{leone2022stabilizerrenyientropy,haug2023stabilizerentropiesand}
\begin{equation}
    M_q\equiv \frac{1}{1-q}\log_2\left[\zeta_q\right],\quad \zeta_q\equiv \sum_{P\in \mathcal{P}_N}\frac{\langle \Psi|P|\Psi \rangle^{2q}}{d},\label{eq:sre}
\end{equation}
through its moments $\zeta_q =  d\int dx\, \Pi(x) x^{2q}$. 
The SRE characterizes the spread of  $|\Psi\rangle\langle \Psi|$ in the Pauli basis, while $M_q$ and $\zeta_q$ are a participation entropy and an inverse participation ratio~\cite{luitz2014shannonrenyientropy,luitz2014universalbehaviorbeyond,sierant2022universalbehaviorbeyond} in the operator space~\cite{haug2023quantifyingnonstabilizernessof,turkeshi2023measuring}.

Computing~\eqref{P} for many-body systems is challenging due to the highly non-local correlations. Previously, only the PS for a few qubit systems or product states have been computed~\cite{Beverland2020}. In the following, we evaluate the Pauli spectrum and the SRE for typical states, and obtain a closed-form expression for Haar random states.

\section{Pauli spectrum of typical quantum states} 
\label{sec:typ}
Our approach is based on the notion of quantum typicality, asserting that quantities of interest evaluated in a statistical ensemble $\mathcal{E} = \{ p(\Psi), |\Psi\rangle\}$ of physical relevance share common, narrowly distributed features~\cite{reimann2007typicality,gemmer2009dynamical}. 
A notable example of quantum typicality is the Eigenstate Thermalization Hypothesis ({ETH})~\cite{deutsch1991quantumstatisticalmechanics,srednicki1994chaosandquantum,Rigol_2008,foini2019eigenstatethermalizationand,foini2019eigenstatethermalizationhypothesis,dymarsky2019newcharacteristicof,brenes2020multipartite,pappalardi2022eigenstatethermalization,pappalardi2023general,pappalardi2023microcanonical,wang2022eigenstatethermalizationhypothesis,dymarsky2022boundoneigenstate,wang2023emergence}, where $\mathcal{E}$ is a microcanonical thermodynamic ensemble, or $k$-designs~\cite{Brando2016,Roberts2017,Ippoliti2022solvablemodelofdeep,cotler2023emergentquantumstate,ippoliti2023dynamicalpurification,fava2023designs}. 
Starting from an ensemble $\mathcal{E}$, for each state $\ket{\Psi}$ from the ensemble, 
and for each Pauli element $P$, we have ${\langle \Psi|P|\Psi\rangle = a_P + R_P(\Psi)}$, where ${a_P = \mathbb{E}_{\mathcal{E}}[\langle\Psi|P|\Psi\rangle]}$  and $R_P(\Psi)$ can be treated as a stochastic variable with zero mean and variance $b_P$. 
In general, $R_P(\Psi)$ is non-Gaussian, as it happens in disordered systems~\cite{luitz2016anomalousthermalization,luitz2016longtail,brenes2020lowfrequency}. Furthermore, $a_P$ and $b_P$ may exhibit complicated dependence on the Pauli string $P$.

Guided by typicality, in many physical situations, we expect that for ${P\neq I}$, 
  ${a_P\approx 0}$, and $R_P(\Psi)$ is  a Gaussian random variable with variance ${b=b_P}$ being the same for all ${P\neq I}$. 
Thanks to normalization condition $\sum_{P\in \mathcal{P}_N} \langle \Psi|P|\Psi\rangle^2/d=1$ this uniquely fixes the PS ``probability'' distribution \eqref{P} for the variable  ${x=\langle \Psi|P|\Psi\rangle}$, 
\begin{equation}
    \Pi_\mathrm{typ}(x) = \left(1-\frac{1}{d^2}\right)  \frac{e^{-x^2/(2b)}}{\sqrt{2 \pi b}}+\frac{1}{d^2} \delta (x-1),\label{eq:pheno}
\end{equation}
where $b=(d+1)^{-1}$ and the delta-function is due to ${P=I}$. 
We expect \eqref{eq:pheno} to hold e.g.~for chaotic systems, up to large-fluctuation corrections~\cite{mondaini2017eigenstatethermalizationin,richter2020eigenstatethermalizationhypothesis,brenes2021outoftime,brenes2020eigenstatethermalizationin}, see below. 

The typicality-based consideration can be extended to  \emph{real} states $|\Psi\rangle$, appearing in time-reversal invariant (TRI)  setups. The TRI  enforces ${\bra{\Psi}P_o \ket{\Psi}=0}$ for $D_o= d(d-1)/2$  Pauli strings $P_o$ containing an odd number of $Y_i$ Pauli matrices. The resulting distribution is 
\begin{equation}
    \Pi_\mathrm{typ}^\mathrm{TRI}(x) = \frac{D_e-1}{d^2}\frac{e^{-x^2/(2b)}}{\sqrt{2 \pi b}}+\frac{1}{d^2} \delta (x-1) + \frac{D_o}{d^2}\delta(x),\label{eq:phenore}
\end{equation}
where $b=(d/2+1)^{-1}$ and $D_e = d^2-D_o$ is the number of Pauli strings $P_e$ with an even number of $Y_i$.

\begin{figure}
    \centering
    \includegraphics[width=\columnwidth]{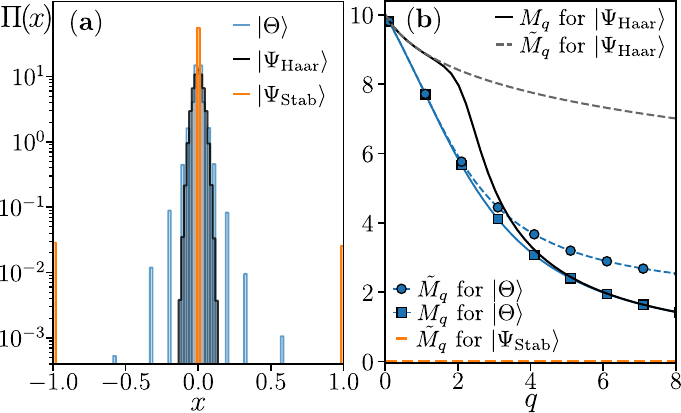}
    \caption{(a) Pauli spectrum for the atypical product state $|\Theta\rangle$ (where we fix $\theta=2 \arctan(\sqrt{2-\sqrt{3}})$ and $\phi=2\arctan(1-\sqrt{2})$), a typical (Haar random) state $|\Psi_\mathrm{Haar}\rangle$, and for a stabilizer state $|\Psi_\mathrm{Stab}\rangle$ generated by a random Clifford unitary for $N=10$ qubits. These distributions manifest distinctive traits, characterizing structural properties of non-stabilizerness. (b) The (filtered) stabilizer entropy ($\tilde{M}_q$) $M_q$. 
    Resolving between $M_q$ for $|\Theta\rangle$ and $|\Psi_\mathrm{Haar}\rangle$ requires precision growing exponentially with $q$, whereas the separation of the $\tilde{M}_q$ values is neat at any $q$. Both 
    $\tilde{M}_q$ and $M_q$ are magic measures and $M_q=\tilde{M}_q=0$ for stabilizer states.}
    \label{fig:1}
\end{figure}

\section{Filtered stabilizer entropy}
\label{sec:filtered}
The distributions~\eqref{eq:pheno} and~\eqref{eq:phenore} imply that the SRE of a typical state is
\begin{equation}
   M_q^\mathrm{typ} = \frac{1}{1-q}\log_2\left[ \frac{(\eta-1) (2b)^q   \Gamma \left(q+\frac{1}{2}\right)}{\sqrt{\pi }d }+\frac{1}{d}\right],\label{eq:gausssre}
\end{equation}
where $\Gamma(x)$ is the gamma function, $b$  is given above and  ${\eta=d^2}$ (${\eta=D_e}$) for complex (real) typical states~\footnote{We note that, since $\Pi(x)$ is exponentially narrow in system size for typical states, the stabilizer entropy $M_q$ is self-averaging.}. 
Leveraging on the interpretation of SRE as the operator participation entropy, we anticipate the system size scaling of SRE as $M_q = D_q N + c_q$, with $D_q$ being the magic density and $c_q$ a sub-leading constant~\cite{haug2023quantifyingnonstabilizernessof}.
From \eqref{eq:gausssre}, $D_q=1/(q-1)$ depends non-trivially on the index $q$ and vanishes in the large $q$ limit. (The subleading term in the  ${N\to\infty}$ limit  is ${c_2=-2}$ (${c_2=-\log_2\, 7}$) and $c_{q>2}=0$ for generic (TRI) typical states.)

We contrast this behavior with the stabilizer entropy for the (atypical) product state $|\Theta\rangle = (\cos(\theta/2)|0\rangle+ e^{-i \phi}\sin(\theta/2)|1\rangle)^{\otimes N}$ for generic $0<\theta<\pi/2$ and $\phi$. 
Although $M_q^\Theta\le M_q^\mathrm{typ}$, we have the magic density $D_q^\Theta = 1/(q-1) + \mathrm{O}(e^{-q})$ for large $q$, implying that the distinction between typical and atypical states is exponentially suppressed in the R\'enyi order $q$, cf. ~Fig.~\ref{fig:1}. 
This counter-intuitive universality of magic density for $q\gg 1$ comes from the identity operator contribution, which, for typical states and large $N$, gives the leading contribution to $\zeta_q$ for any $q>2$. 

Motivated by this, we introduce the \emph{filtered} stabilizer entropy ({FSE})
\begin{equation}
    \tilde{M}_q \equiv \frac{1}{1-q}\log_2\left[\tilde{\zeta}_q\right],\quad \tilde{\zeta}_q\equiv \sum_{P\neq I}\frac{\langle \Psi|P|\Psi \rangle^{2q}}{d-1}\label{eq:tsre},
\end{equation}
where the identity contribution is removed and $\tilde{\zeta}_q$ is normalized such that $\tilde{M}_q = 0$ for the stabilizer states. 
While $\tilde{M}_q$  is functionally dependent on $M_q$, and hence is a magic measure, see Appendix~\ref{app:properties}, 
it readily distinguishes between  typical and atypical quantum many-body states, cf. Fig~\ref{fig:1}. 
Thus, for typical states we find for the magic density $\tilde{M}_q= \tilde{D}_q N + \tilde{c}_q$ in the scaling limit, $\tilde{D}_q=1$ for all $q\geq 1$, while for the product state $\tilde{D}_q^\Theta =D_q^\Theta$. Furthermore, for typical complex (real) states we find $\tilde{c}_q = \log_2( (2q-1)!!)/(1-q)$ [$\tilde{c}_q= \log_2((2q-1)!!)/(1-q)-1$], revealing a rich structure, absent for product states, which have $\tilde{c}_q = 0$ for any $q$. 
In other words, $\tilde{M}_q$ reflects the structural properties of typical states at a \emph{polynomially subleading order} in system size, similarly to the participation entropy quantifying state spread in the computational basis~\cite{mace2019multifractal, backer2019multifractal, sierant2022universalbehaviorbeyond}. 
To further corroborate our findings, we consider the significant class of pseudo-magic atypical states in 
Sec.~\ref{subsec:pseudo}, finding they behave qualitatively similar to the product states discussed here.

\section{Pauli spectrum of Haar random states.}
\label{sec:psHaar}
As our next step, we justify the assumptions behind \eqref{eq:pheno} and \eqref{eq:phenore} for Haar-random states by explicitly evaluating their Pauli spectrum and SRE. 

We first consider the Haar-random real states given by $|\Psi\rangle = O|0\rangle^{\otimes N}$, where $O$ is a real orthogonal matrix taken with Haar measure from the orthogonal group $\mathcal{O}(d)$. The corresponding  Pauli spectrum is given by $\Pi^O_d(x) \equiv \mathbb{E}_O \left[ \sum_{P\in \mathcal P_N} \delta(x- \braket{\Psi|P |\Psi } )/d^2 \right] = \sum_{P\in \mathcal P_N} \mathbb{E}_O \left[ \delta(x- \braket{\Psi|P |\Psi } )/d^2 \right]$, where $\mathbb{E}_O[.]$ denotes the Haar average over the orthogonal group. Due to the reality of $|\Psi\rangle$, $x =\langle \Psi|P|\Psi\rangle = 0$ for $D_o$ Pauli strings which fulfill $P=-P^T$, while for $P=I$ we find $x=1$. These two classes of Pauli strings contribute to Dirac delta peaks in $\Pi^O(x)$ at $x=0$ and $x=1$. The $D_e$ real Pauli strings ($P=P^T$) contribute to the regular part of $\Pi^O(x)$. We note that $\Pi^O_{\mathrm{reg}}(x)=\mathbb{E}_O \left[ \delta(x- \braket{\Psi|P |\Psi } )/d^2 \right]$ is, due to the Haar invariance of $\ket{\Psi}$, the same for each real Pauli string $P$. Hence, it suffices to calculate  $\Pi^O_{\mathrm{reg}}(x)$ for a selected example of Pauli string, which we take as $P=Z_1\equiv Z_1 \otimes I^{\otimes(N-1)}$. Ordering the states $\ket{i}$ of the computational basis $\{ \ket{i} \}$ so that $Z_1\ket{i}=\ket{i}$ for $1\leq i\leq d/2$ and $Z_1\ket{i}=-\ket{i}$ for $d/2 < i\leq d$, and writing $\ket{\Psi}=\sum_i c_i \ket{i}$, where $c_i\in \mathbb{R}$, we find that $x=\sum_{i=1}^{d/2} c_i^2 -\sum_{i=d/2+1}^{d} c_i^2 $. To compute $\Pi^O_{\mathrm{reg}}(x)$, we note that the Haar random state $\ket{\Psi}$ corresponds to a point on a unit sphere $S^{d-1}=\{(c_i)\in \mathbb{R}^{d} : \sum_{i=1}^{d} c_i^2=1\}$ taken with uniform probability. Parameterizing $S^{d-1}$ with polyspherical coordinates~\cite{Vilenkin92}, we find that 
$\ket{\psi} = \cos(\theta)\sum_{i=1}^{d/2} a_i \ket{i} + \sin(\theta)\sum_{i=d/2+1}^{d} a_i \ket{i}$, where the coefficients $a_i$ depend on the remaining angles $\theta_2,\ldots, \theta_d$ of the polyspherical coordinates. This implies that $x=\cos(2\theta)$. Integrating out $\theta_2,\ldots, \theta_d$, the induced measure is $ \Pi^O_{\mathrm{reg}}(\theta) = 2\left(\sin(\theta)\cos(\theta)\right)^{d/2-1}/B(d/4,d/4)$,
see~\cite{Vilenkin92}, 
with $B(x,y)$ being the Beta function. Transforming back the distribution (with the additional Jacobian factor) and taking into account the contributions for $P=I$ and for the $D_o$ Pauli strings with $P^T=-P$, we find
\begin{equation}
\begin{split}
    \Pi_d^\mathrm{O}(x) &=\frac{1}{d^2}\delta(x-1)+ \frac{D_o}{d^2}\delta(x)\\ &\ +\frac{D_e-1}{d^2}\frac{ \left(1-x^2\right)^{d/4-1} \Gamma \left(\frac{1}{4} \left(2+d\right)\right)}{\sqrt{\pi } \Gamma \left(d/4\right)}.
\end{split}
    \label{eq:coeexact}
\end{equation}
A similar argument applies for $|\Psi\rangle = U|0\rangle^{\otimes N}$ with a  Haar-random unitary $U$ from the unitary group $\mathcal{U}(d)$. 
To calculate $ \Pi_d^U(x) \equiv \mathbb{E}_O \left[ \sum_{P\in \mathcal P_N} \delta(x- \braket{\Psi|P |\Psi } )/d^2 \right]$, we proceed in analogously as in the real case above, obtaining $\ket{\Psi} =\sum_{j=1}^d (a_j+i b_j) \ket{j}$, where $a_j, b_j$ are the real and imaginary parts of $\braket{j|\Psi}$. Denoting $c_j=a_j$ and $c_{j+d} = b_j$ for $j\in[1,d]$, the normalization of $\ket{\Psi}$ implies that $\sum_{j=1}^{2d}c_j^2=1$, i.e., the Haar-random state $\ket{\Psi}$ corresponds to a uniformly chosen point on the sphere $S^{2d-1}$. 
Hence, the resulting  distribution is 
given by $\Pi^U_{\mathrm{reg}}(\theta) = 2 (\sin(\theta)\cos(\theta))^{d-1}/B(d/2,d/2)$.
Transforming back with $\theta=\arccos(x)/2$ and considering the $P=I$ term yields
\begin{equation}
\begin{split}
    \Pi_d^\mathrm{U}(x) &=\frac{1}{d^2}\delta(x-1)+ \frac{ d^2-1}{d^2}\frac{ \left(1-x^2\right)^{d/2-1} \Gamma \left(\frac{d+1}{2}\right)}{\sqrt{\pi }  \Gamma \left(d/2\right)}.
\end{split}
    \label{eq:cueexact}
\end{equation}
An alternative derivation of~\eqref{eq:cueexact} making a connection with $k$-design  is delegated to 
Appendix~\ref{app:deriv}.
We note that, in the limit ${N\to\infty}$, \eqref{eq:coeexact} and \eqref{eq:cueexact} converge to typicality-based distributions \eqref{eq:phenore} and \eqref{eq:pheno}, validating our assumptions above. As a consequence, the values of $M_q$ and $\tilde{M}_q$ discussed above accurately capture $q$-dependence of SRE and FSE for Haar random states, for any $N$.

\begin{figure}
    \centering
    \includegraphics[width=\columnwidth]{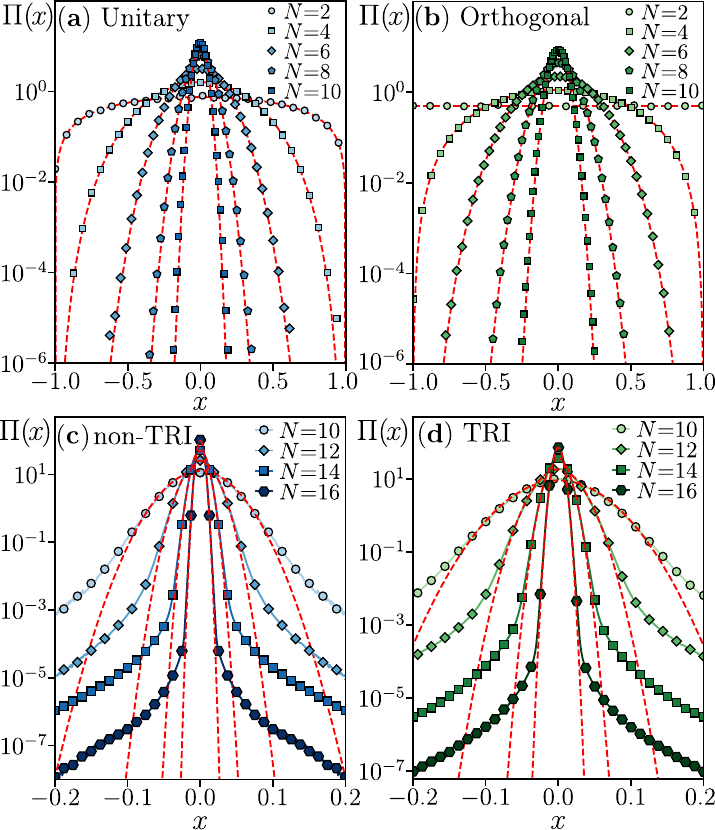}
    \caption{ Pauli spectra of typical states (red dashed lines, Eq.~\eqref{eq:pheno} and ~\eqref{eq:phenore}) and the numerical results for: random unitary circuits (a), random orthogonal circuits (b), mid-spectrum eigenstates of chaotic Hamiltonian $H^\mathrm{sb}$ without TRI, (c), and with TRI $H$, (d), see Text. 
    Only the regular parts of the distributions are plotted. 
    }
    \label{fig:dist}
\end{figure}

\section{Pauli Spectra and filtered stabilizer entropy in many-body settings}
In this section, numerically investigate PS and the FSE in quantum many-body systems to demonstrate the physical relevance of the typicality picture. 

\subsection{Sampling procedure}
Since the Pauli spectrum contains $4^N$ elements, its direct evaluation is only possible for systems comprising a small  ${N\lesssim 12}$ number of qubits. To extend the range of accessible system sizes, we employ the following sampling procedure. To probe state $\ket{\Psi}$, we associate each Pauli string $P$$\neq$$I$ with a probability $\rho( P ) $$ =$$ \langle \Psi|P|\Psi\rangle^2/(d-1)$, and put $\rho(I)$$=$$0$~\footnote{The identity $\sum_{P\in \mathcal{P}_N} \langle \Psi|P|\Psi\rangle^2/d=1$ implies that $\sum_{ P\in \mathcal{P}_N \setminus I}  \rho( P ) = 1$}. We sample the resulting probability distribution over the Pauli strings using the Metropolis-Hastings algorithm~\cite{Metropolis1953equation, Hastings70}. Initialized in a random Pauli string $P_1$, such that $\rho(P_1)>\epsilon$ (we select $\epsilon=10^{-14}$), the algorithm performs $\mathcal N-1 \gg 1$ steps and outputs a sequence of Pauli strings $P_1,\ldots,P_\mathcal{N}$. At the $k$-th step, the algorithm generates a candidate Pauli string $P'_{k+1}$ by multiplying $P_{k}$ by two operators selected with uniform probability from the set $\{ X_1,Z_1,\ldots, X_N, Z_N \}$. The candidate Pauli string is accepted ($P_{k+1} $$=$$ P'_{k+1}$) with probability $p_{\mathrm{acc} }=\min\{ 1, \rho(P'_{k+1})  / \rho(P_{k+1}) \}$, or, otherwise, it is rejected ($P_{k+1} $$=$$ P_{k}$). On average, the algorithm outputs $\mathcal N \rho(P)$ Pauli strings $P$. This allows us to approximate the Pauli spectrum as $\Pi(x) = \sum_{k=1}^{\mathcal N} \rho(P_k)^{-1} \delta( x - \bra{ \Psi } P \ket{\Psi})$, and calculate the FSE~\eqref{eq:tsre} as $\tilde M_q=\log_2( (d-1)^{q-1}\sum_{k=1}^{\mathcal N} \rho(P_k)^{q-1}) /(1-q)$. 
Eliminating the identity Pauli string $I$ (which otherwise would have an anomalously large weight) makes this sampling procedure efficient for typical quantum states that we consider here. In contrast, for a stabilizer state, the $p_{\mathrm{acc} }$ is exponentially small in $N$, and the algorithm is less efficient than the approaches of~\cite{tarabunga2023manybodymagic,lami2023quantum,haug2023stabilizerentropiesand}. In the following, we reach system sizes up to $N=20$ by setting $\mathcal N=10^6$ for $N\leq 16$ and $\mathcal N=10^5$ for $16 < N \leq 20$. 

\subsection{Brick-wall Haar random circuits}
As a first numerical example, we consider brick-wall quantum circuits~\cite{fisher2023randomquantumcircuits} with local two-qubit gates drawn from the Haar distribution on the unitary and orthogonal groups. 
The architecture is the brick-wall circuit, with each unitary layer given by 
\begin{equation}
    U_t = \prod_{i=1}^{N/2}\begin{cases}
        U_{2i-1,2i} & \text{if } t \text{ is even},\\
        U_{2i,2i+1} & \text{otherwise},
    \end{cases}
    \label{eq:cirq}
\end{equation}
and periodic boundary conditions are assumed. 
The unitaries $U_{i,j}$ are drawn from $\mathcal{U}(4)$ ($\mathcal{O}(4)$) for unitary (orthogonal) circuits. We evolve the state $|\Psi_{t_\mathrm{max}}\rangle = U_t| 0 \rangle^{\otimes N}$ up to $t_\mathrm{max}=L$ depth, and compute the average stabilizer entropy with $\mathcal{N}_{\mathrm{real}}=1000$ circuit realizations.

Our results, for circuits of depth $T\sim \mathcal{O}(N)$, demonstrate that the Pauli spectra $\Pi(x)$ in~\eqref{eq:coeexact} and~\eqref{eq:cueexact} are reproduced, see Fig.~\ref{fig:dist}(a,b). 
This fact is expected since random circuits approximate the full Haar distribution being approximate $k$-designs~\cite{Brando2016, Haferkamp2022randomquantum}. 

\subsection{Ergodic quantum many-body systems}
Now, we turn to eigenstates of ergodic quantum systems~\cite{Alessio16rev, abanin2019}, and focus on the Ising model 
\begin{equation}
H_{\mathrm{TFIM}} = \sum_{i=1}^{N}g X_{i} + \sum_{i=1}^{N}h_i Z_{i} + \sum_{i=1}^{N-1} J Z_i Z_{i+1}.
\label{eq:tfim}
\end{equation}
We set open boundary conditions and follow the parameter choice of \cite{Kim13ballistic}: $J$$=$$1$, $g$$=$$(\sqrt{5} + 5)/8$, $h_i$$=$$(\sqrt{5} + 1)/4 $ for $i=2,\ldots, N$ and $h_1 $$=$$-J$. With those specifications, \eqref{eq:tfim} is quantum ergodic, with level statistics adhering to the predictions of Gaussian Orthogonal Ensemble of random matrices~\cite{Meh2004, Haake}. 
We also consider breaking the TRI by $H_1= Y_{L-2} Z_{L-1} +Y_{L-1} Z_{L}$, resulting in the model $H^{\mathrm{sb}} = H_{\mathrm{TFIM}} + H_1$ with level statistics conforming to Gaussian Unitary Ensemble of random matrices. We calculate $n_{\mathrm{ev}} = \min\{d/10,10^4\}$ mid-spectrum eigenstates of $H_{\mathrm{TFIM}}$ and $H^{\mathrm{sb}}$ with POLFED algorithm~\cite{Sierant20POLFED, Sierant23Stability} for $10 \leq N \leq 20$, perform the sampling of Pauli strings with the Metropolis-Hastings algorithm, and average the results over the eigenstates.

\begin{figure}
    \centering
    \includegraphics[width=0.95\columnwidth]{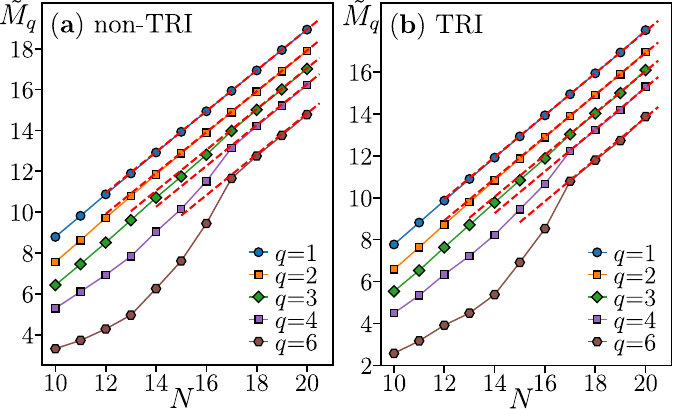}
    \caption{ The filtered stabilizer entropy $\tilde M_q$ at  $q=1,\dots,6$ for the random Haar states (red dashed lines) and eigenstates of chaotic Hamiltonian for systems (markers) without (a) and with (b) time-reversal invariance. 
    The data for distinct $q$ are shifted downwards by $(q-1)/2$ for clarity. }
    \label{fig:3}
\end{figure}

We obtain quantitatively similar Pauli spectra $\Pi(x)$ for both $H_{\mathrm{TFIM}}$ and $H^{\mathrm{sb}}$, as shown in Fig.~\ref{fig:dist}. The spectra follow the Haar state predictions \eqref{eq:cueexact} and \eqref{eq:coeexact}, respectively, for the systems without and with TRI, down to a tail, which decays exponentially with $x$, and with the weight being exponentially suppressed in $N$. 
The Gaussian core of $\Pi(x)$ around zero indicates that the expectation value of Pauli strings in chaotic eigenstates tends to be exponentially close to zero as a function of $N$. This trend is prevalent among the vast majority of Pauli strings, which are typically non-local operators, and hence, their description is beyond the standard formulation of ETH, cf.~\cite{khaymovich2019eigenstatethermalizationrandom, Lydzba21single, Ulcakar22}. The tails of $\Pi(x)$ reflect a finer structure encoded in eigenstates of the Hamiltonian. Local Pauli strings, such as $P=X_i$, follow the ETH, exhibiting non-zero thermal expectation value $f_P(E)$ varying smoothly with energy $E$, up to fluctuations suppressed exponentially in $N$~\cite{Beugeling14finite, Ikeda13finite, PhysRevE.97.012140, Mierzejewski20}. While instances of such local operators are exponentially rare in $\mathcal P_N$, they constitute one of the contributions to the tails of $\Pi(x)$; see 
Appendix~\ref{app:additional} for further discussions. The presence of tails in $\Pi(x)$ is also visible in the behavior of the FSE shown in Fig.~\ref{fig:3}. With  $q$ increasing, the tail affects the value of  $\tilde M_q$ more, shifting it below the value of $\tilde{M}_q^\mathrm{typ}$~\eqref{eq:gausssre}. 
However, due to the suppression of tails for large $N$, we again recover the typical scaling 
of FSE, $\tilde M_q= N + \tilde c_q$, for all considered values of $q$ for the mid-spectrum eigenstates of both $H_{\mathrm{TFIM}}$ and $H^{\mathrm{sb}}$.

\begin{figure}
    \centering
    \includegraphics[width=\columnwidth]{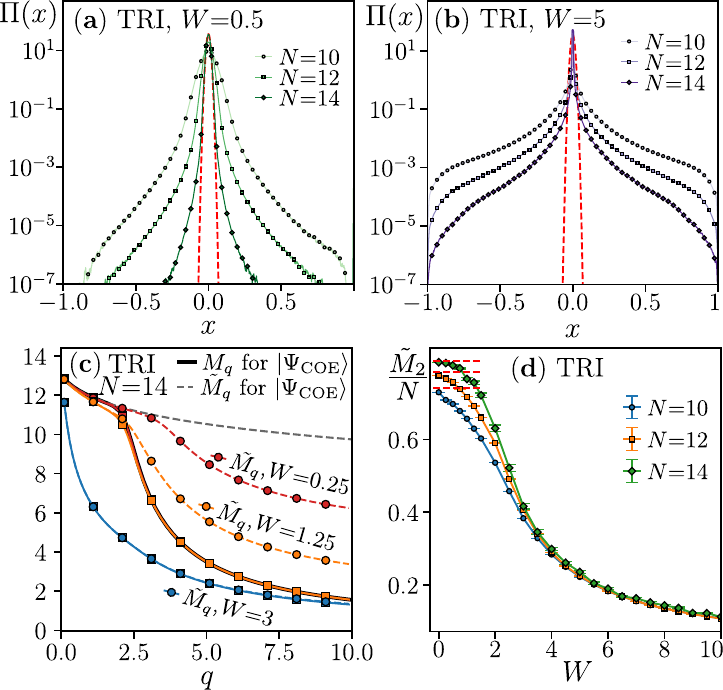}
    \caption{Ergodicity breaking in disordered Ising model. Pauli spectrum $\Pi(x)$ in the ergodic (a), and MBL (b) regime for system size $N=10,12,14$ compared with the  analytical prediction $\Pi_d^\mathrm{O}(x)$ for $d=2^{14}$. (b) The (filtered) stabilizer entropy ($\tilde{M}_q$) $M_q$ for the typical state $\ket{\Psi_{O}}$, and mid-spectrum eigenstates of $H_{\mathrm{dis}}$ in the ergodic regime ($W=0.25, 1.25$) and in MBL regime ($W=3$) shown as function of the R\'{e}nyi index $q$. (c) The density of filtered stabilizer entropy $\tilde{M}_q/N$ as a function of disorder strength $W$, the red dashed lines denote the result for Haar-random real states.}
    \label{fig:end2}
\end{figure}

\subsection{Ergodicity breaking}
The Ising model with Hamiltonian $H_{\mathrm{TFIM}}$, \eqref{eq:tfim}, is a paradigmatic quantum ergodic system following the ETH~\cite{Alessio16rev}. The ergodicity of the system may be broken via the phenomenon of many-body localization (MBL) induced by a sufficiently strong on-site disorder potential~\cite{abanin2019, sierant2024mbl}. The Hamiltonian of the disordered system reads
\begin{equation}
    H_{\mathrm{dis}}= H_{\mathrm{TFIM}} + \sum_{i=1}^N w_i Z_i, 
\end{equation}
where the on-site fields $w_i$ are taken as independent random variables drawn uniformly from the interval $[-W, W]$ with $W$ being the disorder strength. We calculate $n=100$ mid-spectrum eigenstates of $H_{\mathrm{dis}}$ and average the results over more than $100$ disorder realizations. Tuning $W$, we interpolate between the ergodic regime at weak disorder and an MBL regime at $W\gg 1$.

In the ergodic regime, for $W=0.5$, the Pauli spectrum $\Pi(x)$ contains a Gaussian peak characteristic for the real Haar random states~\eqref{eq:coeexact} and an exponential tail at larger $|x|$, see Fig.~\ref{fig:end2}~(a), analogously to the clean system, $W=0$. The increase of $W$ changes the structure of mid-spectrum eigenstates, which cease to resemble the typical states. The Pauli spectrum at $W=5$ has a narrow peak close to $x=0$ and possesses tails at large $|x|$, as shown in Fig.~\ref{fig:end2}~(b). Further growth of $W$ raises the weight of the central peak and decreases the weights of the large $|x|$ tails. In the $W\gg1$ limit, the eigenstates of $H_{\mathrm{dis}}$ are eigenstates of $Z_j$ operators and, consequently, stabilizer states with Pauli spectrum localized at $x \in \{-1,0,1\}$. 

The filtered stabilizer entropy $\tilde{M}_q$ clearly distinguishes the eigenstates of $H_{\mathrm{dis}}$ within the ergodic regime at larger values of $q$, as exemplified for $W=0.25$ and $W=1.25$ in Fig.~\ref{fig:end2}~(c), in contrast to the $M_q$ curves which nearly overlap for $q\gtrsim 2.5$. At the same time, $\tilde{M}_q$, even in a weakly disordered system, $W=0.25$, deviates from the real Haar random state results at larger $q$, consistently with the more pronounced role played by the exponential tail at larger $q$. Finally, the difference between stabilizer entropy $M_q$ for ergodic eigenstates ($W=0.25$) and non-ergodic eigenstate ($W=3$) vanishes in the large $q$ limit, a trend that is not occurring for the filtered stabilizer entropy $\tilde{M}_q$.

Plotting the filtered stabilizer entropy as a function of disorder strength $W$ in Fig.~\ref{fig:end2}~(d), we observe that the magic density $\tilde{M}_q/N$ may play a role of ergodicity-breaking indicator, distinguishing the ergodic regime at small $W$ in which $\tilde{M}_q/N$ tends to the typical state result and the MBL regime in which $\tilde{M}_q/N$ becomes a system size independent constant smaller than unity. The behavior of magic density is similar to the multifractal dimension characterizing the spread of many-body wave-function in a fixed basis of Hilbert space~\cite{mace2019multifractal}.

\subsection{Pseudomagic}
\label{subsec:pseudo}
To contextualize the scope of our work, in the following we compare the Pauli spectra and the (filtered) stabilizer entropy of typical quantum states with results for ensembles of \textit{pseudomagic} states~\cite{Gu24pseudomagic}. 

Discriminating quantum states is a crucial task in quantum simulation and computation. We say two ensembles of states $\{\ket{\Phi_k}\}$ and $\{ \ket{\Psi_k}\} $ on $N$ qubits are \emph{computationally indistinguishable} if no quantum algorithm with a polynomial number $p(N)$ of elementary operations can differentiate between the ensembles $\rho= \mathbb{E}_k \left[ \ket{\Psi_k}\bra{\Psi_k}^{\otimes p(N)}\right]$ and $\sigma= \mathbb{E}_k \left[ \ket{\Phi_k}\bra{\Phi_k}^{\otimes p(N)}\right]$ with more than negligible probability~\cite{Aaronson23pseudoEnt,cheng2024pseudoentanglementtensornetworks,feng2024dynamicspseudoentanglement}.
Pseudomagic states $\ket{\Phi_k}$ are characterized by small non-stabilizerness while being computational indistinguishable from quantum states $\ket{\Psi_k}$ with rich magic resources. 
Subspace phase states (SPS), defined as
\begin{equation}
    \ket{SPS}\equiv \frac{1}{\sqrt{|S|}}\sum_{\sigma \in S}(-1)^{f(\sigma )}\ket{\sigma },
    \label{eq:subsetphasestates}
\end{equation}
where $f:\{0,1\}^N\to \{0,1\}$ is a pseudorandom function and $S$ is a subset $S\subseteq \{0,1\}^N$ containing $|S|=2^k$ elements, are computationally indistinguishable from Haar random states~\cite{Aaronson23pseudoEnt}. Moreover, the stabilizer entropies of SPS are $M_q(\ket{SPS}) = O(k)$~\cite{Gu24pseudomagic}. Hence, for fixed $k$, the SPS states have high-pseudomagic resources.

\begin{figure}
    \centering
    \includegraphics[width=\columnwidth]{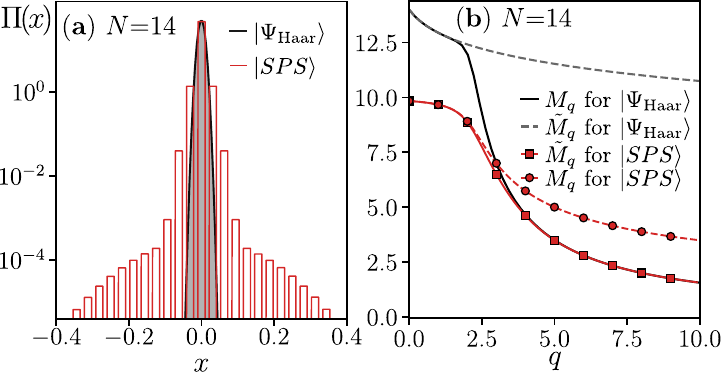}
    \caption{Non-stabilizerness in sparse phase states (SPS). (a) Pauli spectrum for SPS state $\ket{SPS}$ with $|S|=64$ for system size $N=14$ is compared with the typical state result $\Pi^{U}_d(x)$ for $d=2^{14}$. (b) The (filtered) stabilizer entropy ($\tilde{M}_q$) $M_q$ for the typical state $\ket{\Psi_{\mathrm{Haar}}}$ and SPS shown as a function of the R\'{e}nyi index $q$. }
    \label{fig:end1}
\end{figure}

In the following, we contrast the Pauli spectra and filtered stabilizer entropy of Haar random states $\ket{\Psi_{\mathrm{Haar}}}$, i.e., the states with high magic with the properties of SPS. We take $f$ to be equal to $0$ ($1$) with probability $0.5$, and fix the size of the subset $S$ as $|S|=2^6$ and the qubit number as $N=14$. The Pauli spectrum $\Pi(x)$, averaged over $100$ realizations of SPS, shown in Fig.~\ref{fig:end1}~(a), contains discrete peaks which form a central maximum and a tail at $|x|>0.1$. Inspecting the Pauli spectrum allows for an immediate distinction between the high-pseudomagic SPS and the typical Haar random states. This is consistent with the computational indistinguishability of the SPS and Haar random states since calculating the Pauli spectrum requires computation of \text{exponentially} many in $N$ expectation values of Pauli strings. The stabilizer entropy $M_q$ for $q \lesssim 2$ neatly separates the SPS from the Haar random states, see Fig.~\ref{fig:end1}~(b). However, the difference between the stabilizer entropy $M_q$ for SPS and $\ket{\Psi_{\mathrm{Haar}}}$ is vanishing with the increase of $q$. In contrast, the filtered stabilizer entropy, denoted by dashed lines in Fig.~\ref{fig:end1}~(b), clearly distinguishes between SPS and Haar random state for arbitrary $q$.

The results for highly pseudomagical SPS qualitatively follow the findings for product states described in 
%the Main Text
Sec.~\ref{sec:psHaar}. A reasoning for SPS analogous to \cite{Gu24pseudomagic} shows that $2\log_2(|S|) =2k \ge \tilde{M}_0 \ge \tilde{M}_q$ for any $q>0$. Hence,  the gap between the Haar state $\tilde{M}_q(\ket{\Psi_{\mathrm{Haar}}}) = N - \tilde{c}_q$ and the SPS, for which $\tilde{M_q}(\ket{SPS}) = O(k)$ is significant at any R\'{e}nyi index $q$, showing the utility the filtered stabilizer entropy.

\section{Discussion and conclusion.}
We have demonstrated that the Pauli spectrum unveils the magic structural properties of many-qubit states. We proposed a phenomenological picture of the Pauli spectrum based on quantum typicality, which we corroborated by deriving the exact expressions for random Haar states. Observing that the identity operator contribution dominates the value of stabilizer entropies for R\'enyi index $q>2$, we have introduced the filtered stabilizer entropy~\eqref{eq:tsre}. This magic measure  readily discriminates 
the typical states from the atypical ones, such as product states. Our extensive numerical simulations show that physically relevant states, such as those prepared by random circuits or mid-spectrum eigenstates of ergodic Hamiltonians, adhere to predictions of the typicality-based picture of the Pauli spectrum exponentially fast with the system size. Notably, the mid-spectrum eigenstates of chaotic Hamiltonians differ from the random circuit states and the Haar random states at the level of the Pauli spectrum by the presence of exponentially suppressed tails, which reveal richer structures encoded in the Hamiltonian.

We find that the filtered magic density is maximal, i.e., $\tilde{D}_q=1$, for typical states at any Renyi index $q$, demonstrating the challenges to prepare such states in digital quantum devices~\cite{Preskill_2018}.
The Pauli spectrum encompasses the stabilizer entropies, enabling a more complete characterization of non-stabilizerness. Moreover, its adherence to results for typical states signals the ergodicity of the system. 
The results presented for the ensemble of SPS states with high-pseudomagic, and for the disordered Ising model show how our results for typical states derived in this work can be employed as a starting point for analysis of magic resources in more complicated scenarios. These considerations can be naturally extended to other types of ergodicity-breaking phenomena such as quantum scars~\cite{Serbyn2021}, ergodicity breaking in deterministic potentials~\cite{Yao21}, or due to gauge invariance~\cite{Brenes18}. Moreover, our results naturally correspond to a long evolution time limit for quantum quenches in ergodic many-body systems, as shown for random quantum circuits~\cite{turkeshi2024spreading,tirrito2024anticoncentrationmagicspreadingergodic} and doped Clifford circuits~\cite{haug2024saturation}.

Our results can be generalized to qudits, but for odd and prime local Hilbert space dimensions, the magic monotones~\cite{gross2006hudsonstheoremfor,Gross2006,howard2017application,white2021CFT,white2020mana} and their efficient sampling~\cite{tarabunga2023critical} have been demonstrated. Constructing magic monotones for qubits is still an outstanding problem. A high-end goal left for future work is building such magic monotones in terms of the Pauli spectrum -- a task facilitated by the exact results presented here. Magic phase transitions, driven by competition of Clifford and non-Clifford resources~\cite{leone2023phase,niroula2023phase,Turkeshi2024,bejan2023dynamical,fux2023entanglementmagic}, constitute an interesting application of our results.

\begin{acknowledgments}
We thank M. Dalmonte, R. Fazio, G. Fux, M. Lewenstein, L. Piroli, S. Pappalardi, P. S. Tarabunga, E. Tirrito, and L. Vidmar for insightful discussions. 
X.T. acknowledge DFG under Germany's Excellence Strategy – Cluster of Excellence Matter and Light for Quantum Computing (ML4Q) EXC 2004/1 – 390534769, and DFG Collaborative Research Center (CRC) 183 Project No. 277101999 - project B01. 
A.\,D.\ is supported by the National Science Foundation under Grant No.~PHY 2310426.
P.S. acknowledges support from: ERC AdG NOQIA; MICIN/AEI (PGC2018-0910.13039/501100011033, CEX2019-000910-S/10.13039/501100011033, Plan National FIDEUA PID2019-106901GB-I00, FPI; MICIIN with funding from European Union NextGenerationEU (PRTR-C17.I1): QUANTERA MAQS PCI2019-111828-2); MCIN/AEI/ 10.13039/501100011033 and by the "European Union NextGeneration EU/PRTR" QUANTERA DYNAMITE PCI2022-132919 within the QuantERA II Programme that has received funding from the European Union's Horizon 2020 research and innovation programme under Grant Agreement No 101017733Proyectos de I+D+I "Retos Colaboración" QUSPIN RTC2019-007196-7); Fundació Cellex; Fundació Mir-Puig; Generalitat de Catalunya (European Social Fund FEDER and CERCA program, AGAUR Grant No. 2021 SGR 01452, QuantumCAT \ U16-011424, co-funded by ERDF Operational Program of Catalonia 2014-2020); Barcelona Supercomputing Center MareNostrum (FI-2023-1-0013); EU (PASQuanS2.1, 101113690); EU Horizon 2020 FET-OPEN OPTOlogic (Grant No 899794); EU Horizon Europe Program (Grant Agreement 101080086 — NeQST), National Science Centre, Poland (Symfonia Grant No. 2016/20/W/ST4/00314); ICFO Internal "QuantumGaudi" project; European Union's Horizon 2020 research and innovation program under the Marie-Skłodowska-Curie grant agreement No 101029393 (STREDCH) and No 847648 ("La Caixa" Junior Leaders fellowships ID100010434: LCF/BQ/PI19/11690013, LCF/BQ/PI20/11760031, LCF/BQ/PR20/11770012, LCF/BQ/PR21/11840013). Views and opinions expressed are, however, those of the author(s) only and do not necessarily reflect those of the European Union, European Commission, European Climate, Infrastructure and Environment Executive Agency (CINEA), nor any other granting authority. Neither the European Union nor any granting authority can be held responsible for them.

\textbf{Data availability.}
The numerical data for this work are given in Ref.~\cite{turkeshi_2025_14696597}. 

\end{acknowledgments}

\appendix 

\section{Combinatoric derivation of typical stabilizer R\'enyi entropy and Pauli spectrum for Haar states.}
\label{app:deriv}
In order to obtain the typical stabilizer R\'enyi entropy and the Pauli spectrum, we begin computing the Haar average $\overline{\zeta}_q\equiv \mathbb{E}_{U}[\zeta_q(U|0\rangle^{\otimes N})]$, where $\mathbb{E}_U [.]$ denotes
the Haar average over the unitary group, see also~\cite{turkeshi2023errorresilience}.
Recalling that $d=2^N$ is the Hilbert space dimension, the computation is recast in terms of $2q$ replicas
\begin{equation}
    \overline{\zeta_q} = 
    \mathbb{E}_{U}\sum_{P\in \mathcal{P}_N}\frac{1}{d}\mathrm{Tr}\left(U^{\otimes (2q)} P^{\otimes (2q)} (U^{\dagger})^{\otimes (2q)}|\Psi\rangle\langle\Psi|^{\otimes (2q)}\right).
\end{equation}
Because of the linearity of the trace and the Haar average, we can exchange the order of the operations. The Haar average over the unitary group relies on the Schur-Weyl duality~\cite{Roberts2017} and leads to
\begin{equation}
    \overline{\zeta_q} = \sum_{\sigma \in S_{2q}} b_\sigma \mathrm{Tr}\left( T_\sigma |\Psi\rangle\langle \Psi|^{\otimes (2q)} \right) =  \sum_{\sigma \in S_{2q}} b_\sigma
\end{equation}
where the coefficients 
\begin{equation}
    b_\sigma =\sum_{P\in \mathcal{P}_N} \sum_{\tau\in S_{2q}}W_{\sigma,\tau} \mathrm{Tr}( P^{(2q)} T_\tau)/d
\end{equation}
have to be evaluated, and $W_{\sigma,\tau}$ is the Weingarten matrix~\cite{Collins2006,turkeshi2023errorresilience}. Since each Pauli string is given by $P=P_1\otimes P_2\otimes \cdots \otimes P_N$, with $P_i$ Pauli matrices acting on the $i$-th site, it follows that 
\begin{equation}
    b_\sigma = \sum_{\tau\in S_{2q}} W_{\sigma,\tau}\mathrm{Tr}\left(\Lambda_q^{\otimes N} T_\tau\right)
\end{equation}
where the $\Lambda_q = \sum_{P_i=I_i,X_i,Y_i,Z_i} P^{\otimes(2q)}/2$ emerge from resummation of the local Pauli matrices. Furthermore, since $T_\tau = t_\tau^{\otimes N}$ acts independently on different sites~\cite{Gross2021}, we have $b_\sigma =\sum_{\tau\in S_{2q}} W_{\sigma,\tau} \mathrm{Tr}(\Lambda_q t_\tau)^N$. 
The form of $\Lambda_q$ implies that all permutations with the same cycle structure contribute equally. Furthermore, since $P_i^2 = I_i$, $\delta_{P_i,I_i}$ appears if any cycle has a odd number of elements. Collecting these facts, we have 
\begin{eqnarray}
    \overline{\zeta_q} =& \sum_{\sigma,\tau \in S_{2q} }  W_{\sigma,\tau} \mathrm{Tr}(\Lambda_q t_\tau)^N = \quad\quad \quad 
   \\ 
   = \frac{1}{\prod_{i=0}^{q-1} (d+i)} & \sum_{\lambda \vdash 2q} d_\lambda \begin{cases}
        2^{\mathrm{len}(\lambda)-1} & \text{if $\lambda \in $ E, 
        }\\ 
        2^{\mathrm{len}(\lambda)+1} & \text{otherwise,} \nonumber
    \end{cases} 
\end{eqnarray}
where $\lambda \in$ E if and only if any cycle $r\in \lambda$ contains odd number of elements.
In the above steps, $\lambda\vdash 2q$ are the cycle structures over $2q$ elements and $d_\lambda$ is the number of permutations whose cycle structure is the same of $\lambda$~\cite{turkeshi2023measuring}, while  $\mathrm{len}(\lambda)$ is the number of cycles in $\lambda$. 
The sum can be performed analytically, and gives
\begin{equation}
    \overline{\zeta}_q = \frac{1}{d} \left(1+\frac{2 \left(d-1\right) \Gamma \left(q+\frac{1}{2}\right) \Gamma \left(\frac{3+d}{2} \right)}{\sqrt{\pi } \Gamma \left(\frac{2q+d+1}{2}\right)}\right).\label{eq:haarzeta}
\end{equation}
From~\eqref{eq:haarzeta} we can already compute $M_q$ and $\tilde{M}_q$, which are self-averaging~\footnote{Computing $\mathrm{std}(\zeta_q) = \sqrt{\overline{\zeta_q^2}-\overline{\zeta_q}^2}$ one finds $\mathrm{std}(\zeta_q)/\overline{\zeta_q}\sim \mathcal{O}(\exp(-N))$.}
\begin{align}
        M_q^\mathrm{typ} &= \frac{1}{1-q} \log_2 \left[\frac{1}{d}+\frac{2 \left(d-1\right) \Gamma \left(q+\frac{1}{2}\right) \Gamma \left(\frac{3+d}{2} \right)}{d \sqrt{\pi } \Gamma \left(\frac{2q+d+1}{2}\right)}\right],\\
    \tilde{M}_q^\mathrm{typ} &= \frac{1}{1-q} \log_2 \left[\frac{2 \Gamma \left(q+\frac{1}{2}\right) \Gamma \left(\frac{3+d}{2} \right)}{\sqrt{\pi } \Gamma \left(\frac{2q+d+1}{2}\right)}\right].\label{eq:zioA}
\end{align}
Next, we introduce the generating function~\cite{calabrese2008spectrum}
\begin{equation}
    f(z) = \sum_{q=1}^N \frac{1}{d}\overline{\zeta}_q z^{-q} = d\int dx \frac{x^2 \Pi(x)}{z-x^2}\label{eq:cahcy}
\end{equation}
where we used the definition of the Pauli spectrum $\Pi(x)$. Resumming the series, we find 
\begin{equation}
    f(z) =  \frac{d^{-1}}{z-1}+\frac{(d-1)}{dz}{_2F_1\left(1,\frac{3}{2};\frac{1}{2} \left(3+d\right);\frac{1}{z}\right)}.
\end{equation}
Lastly $\Pi(x)$ is obtained via Cauchy integration~\eqref{eq:cahcy}, leading to the Haar Pauli spectrum
\begin{equation}
    \Pi_d^\mathrm{U}(x) = \frac{d^2-1}{d^2}\frac{ \left(1-x^2\right)^{-1+d/2} \Gamma \left(\frac{1}{2} \left(1+d\right)\right)}{\sqrt{\pi } \Gamma \left(d/2\right)} + \frac{1}{d^2}\delta(x-1), \label{eq:haarpx}
\end{equation}
recasting the result in 
Sec.~\ref{sec:psHaar}. 
Similar considerations apply also for the time reversal invariant case (TRI). In that case, the Schur-Weyl duality relates moments of the orthogonal Haar group to the Brauer algebra of pairings $\mathfrak{B}(d)$~\cite{Molev2012,Collins2006}. 
We will not pursue here these more intricate computations, and 
instead report the typical $M_q$ and $\tilde{M}_q$, calculated from the exact form of the Pauli spectrum for the TRI case shown in 
Sec.~\ref{sec:psHaar}, and given respectively by 
\begin{align}
    M_{q,\mathrm{TRI}}^\mathrm{typ} &= \frac{1}{1-q}\log_2\left[\frac{2 (d-1) \Gamma \left(\frac{d+6}{4}\right) \Gamma \left(\frac{2q+1}{2}\right)}{d\sqrt{\pi } \Gamma \left(\frac{d+2+4q}{4}\right)}+\frac{1}{d}\right],\\
    \tilde{M}_{q,\mathrm{TRI}}^\mathrm{typ} &= \frac{1}{1-q}\log_2\left[\frac{2  \Gamma \left(\frac{d+6}{4}\right) \Gamma \left(q+\frac{1}{2}\right)}{\sqrt{\pi } \Gamma \left(\frac{d+2+4q}{4}\right)}\right].\label{eq:zioB}
\end{align}
In the scaling limit, these results converge to the phenomenological values of SRE and FSE discussed in 
Sec.~\ref{sec:typ}.

\section{Properties of the filtered stabilizer entropy }
\label{app:properties}
From its definition, it is clear that the filtered stabilizer entropy is related to the stabilizer entropy. Indeed, we have, cf. 
Sec.~\ref{sec:filtered},
\begin{equation}
    \tilde{\zeta}_q = \frac{d \zeta_q - 1}{d-1}. 
\end{equation}
It follows that (I) $\tilde{\zeta}_q(|\Psi\rangle)=1$ (hence $\tilde{M}_q=0$) for all $q$ iff $|\Psi\rangle$ is a stabilizer state, since $\zeta_q=1$ for all $q$ iff $|\Psi\rangle$ is a stabilizer state.
Furthermore (II) $\tilde{\zeta}_q$ (hence $\tilde{M}_q$) is invariant under Clifford conjugation, namely $\tilde{\zeta}_q(C|\Psi\rangle) = \tilde{\zeta}_q(|\Psi\rangle)$ for any Clifford unitary $C\in \mathcal{C}_N$. This follows from the definition of $\tilde{\zeta}_q$ and the fact that Pauli strings are mapped to single Pauli strings by Clifford unitaries. 
Lastly, (III) $\tilde{M}_q$ is subadditive. Consider a product state $|\Psi\rangle = |\psi_A\rangle\otimes |\psi_B\rangle$, with $|\psi_A\rangle$ leaving in the $d_A=2^{N_A}$ dimensional Hilbert space of $N_A$ qubits, and $|\psi_B\rangle$ on the $d_B = d/d_A = 2^{N_B}$ complementary space on $N_B=N-N_A$ qubits. It follows that
\begin{equation}
\begin{split}
    \tilde{\zeta}_q(|\Psi\rangle) 
    &= \frac{(d_A-1)d_B}{d-1}\tilde{\zeta}_q(|\psi_A\rangle) \zeta_q(|\psi_B\rangle)\\ 
    &\ge \tilde{\zeta}_q(|\psi_A\rangle) \tilde{\zeta}_q(|\psi_B\rangle)
\end{split}
\end{equation}
and in particular $\tilde{M}_q(|\Psi\rangle) \le \tilde{M}_q(|\psi_A\rangle)+\tilde{M}_q(|\psi_B\rangle)$. Furthermore, in the scaling limit $N_A,N_B\to\infty$, the inequality is saturated. Hence, (IIIbis) $\tilde{M}_q$ is asymptotically additive. 
Nevertheless, $\tilde{M}_q$ is non-monotone. The proof follows from the explicit counterexamples in~\cite{haug2023stabilizerentropiesand}, 
which show that the FSE $\tilde{M}_q$ fails to be monote in the same way as the SRE $M_q$. 

\section{Additional numerical results}
\label{app:additional}
\begin{figure}
    \centering
    \includegraphics[width=0.97\columnwidth]{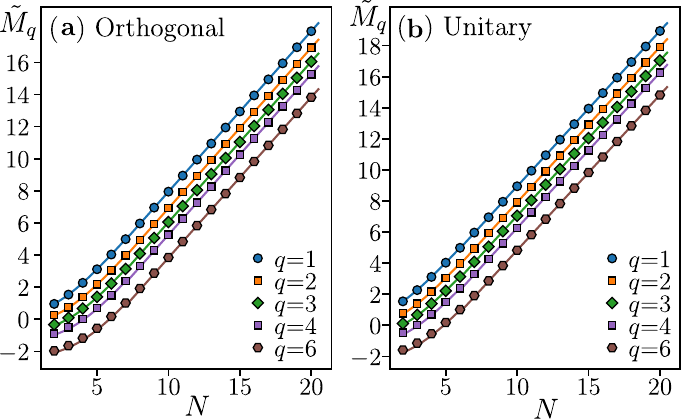}
    \caption{Exact formulas for the FSE of Haar random states (solid lines) vs numerical data (points) for orthogonal (a) and unitary (b) ensembles. The FSE are shifted by $(q-1)/2$ for clarity of the plots.}
    \label{fig:sup1}
\end{figure}

For completeness, we include a numerical test of random circuits generated by unitary and orthogonal two-body gates, with the architecture specified in~\eqref{eq:cirq}. 
We evolve the state $|\Psi_{t_\mathrm{max}}\rangle = U^t| 0 \rangle^{\otimes N}$ up to $t_\mathrm{max}=L$ depth, and compute the average stabilizer entropy with $\mathcal{N}_{\mathrm{real}}=1000$ disorder realizations. The results are summarized in Fig.~\ref{fig:sup1}. We find that, up to the statistical errors associated with the number of circuit realizations and sampling of FSE with the Monte Carlo method (at $N > 12$), the exact Haar predictions match the finite depth circuit numerics, providing a strong benchmark of our analytical arguments. In the same fashion, the average $\tilde{M}_q$ match the exact formulae~\eqref{eq:zioB} and~\eqref{eq:zioA} for TRI (Fig.~\ref{fig:sup1}(b) ) and non-TRI (Fig.~\ref{fig:sup1}(d)) systems, respectively.

\begin{figure}
    \centering
    \includegraphics[width=0.75\columnwidth]{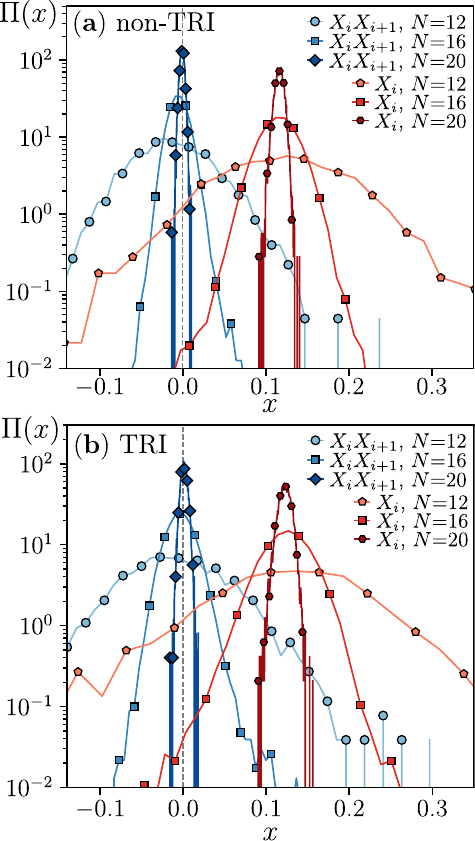}
    \caption{Local structure in the Pauli spectrum in mid-spectrum eigenstates of the Ising model. The Pauli strings $\sigma = X_i X_{i+1}$ (we consider $i=3,\ldots N-3$ to diminish the boundary effects) are distributed typically, according to a Gaussian distribution with standard deviation $\propto 2^{-L}$, decaying exponentially with the system size $L$, and with mean that converges rapidly to $0$. In contrast, the Pauli strings $P = X_i $ (for $i=3,\ldots N-3$) are distributed according to a Gaussian distribution with standard deviation $\propto 2^{-L}$, but with mean that converges to $x \neq 0$. The results are qualitatively the same for the model with TRI symmetry broken, as shown in panel (\textbf{a}), and in the TRI model, as shown in panel (\textbf{b}).}
    \label{fig:sup2}
\end{figure}

Finally, we identify one of the contributions to the tails of the Puali spectra of chaotic eigenstates. The ETH ansatz, typically formulated for a local observable $O$ and an eigenstate $\ket{\Psi}$ of an ergodic Hamiltonian $H$, (with $H \ket{\Psi} = E \ket{\Psi}$) implies that 
\begin{equation}
    \bra{\Psi } O \ket{\Psi} = f_O(E) + \Sigma_O(E),
    \label{eq:ethd}
\end{equation}
where $f_O(E)$ is a smooth function of energy $E$, and $\Sigma_O(E)$ represents fluctuations that are suppressed exponentially with $N$. In the following, we discuss how \eqref{eq:ethd} is reflected in the Pauli spectra $\Pi(x)$ of the chaotic eigenstates.

The Pauli spectra shown in 
Fig.~\ref{fig:dist} consist of a well-pronounced Gaussian core and the exponentially suppressed tail. This implies that the fraction of the Pauli strings $P$ that satisfy equation \eqref{eq:ethd} with $f_P(E)=0$, at the probed energies $E$ close to the middle of the spectrum of the considered Ising models, is increasing, towards unity, with $N$. Moreover, the fluctuations $\Sigma_O(E)$ for such operators are decaying exponentially with $N$. Notably, these Pauli strings are typically highly non-local operators. 

The local structure of the mid-spectrum eigenstates of Ising model is reflected in properties of Pauli strings $P_{\mathrm{loc}}$ that are \emph{local operators}. Since the Ising model is ergodic, the ETH ansatz \eqref{eq:ethd} applies with $O = P_{\mathrm{loc}}$. Importantly, for certain types of local Pauli string, we find that $f_{  P_{\mathrm{loc}} }(E) \neq 0$, and that $f_{  P_{\mathrm{loc}} }(E)$ does not converge to $0$ with increasing $N$. The ratio of the number of the local Pauli strings to the total number of Pauli strings included in $\mathcal P_N$ is, however, exponentially small in $N$. This results in one contribution to the exponential tails of the Pauli spectra observed in Fig.~\ref{fig:dist}. 
To provide concrete examples, we have considered $P_{\mathrm{loc}}$ to be $X_i$, $X_i X_{i+1}$, $Z_i$, or $Z_i Z_{i+1}$ and haven chosen $i$ away from the boundaries of the system. For operators $P_{\mathrm{loc}} = X_i, Z_i, Z_i Z_{i+1}$ we have found that $f_{  P_{\mathrm{loc}} }(E) \neq 0$ (in passing, we note that each of those operators appears in the Hamiltonian of the Ising model). In contrast, certain local operators also exhibit $f_{  P_{\mathrm{loc}} }(E)=0$, an example of such an operator is $X_i X_{i+1}$. The described properties are illustrated in Fig.~\ref{fig:sup2}. 

The above considerations emphasize the role of the exponential tails as a feature the Pauli spectra that reflects the fine structure encoded in eigenstates of ergodic systems. While we have identified one mechanism which contributes to the exponential tails of the Pauli spectra, we have also verified that there exist rare non-local operators with atypically large expectations values which also contribute to the tails of the spectra. Understanding of the properties of such Pauli strings is left for future work. Notably, the exponential tails are absent in the Pauli spectra for the orthogonal and unitary circuits implying total scrambling of the local quantum information at the considered circuit depths.

\bibliography{newbib}
\bibliographystyle{apsrev4-2}

\end{document}